\documentclass[12pt]{article}
\usepackage{graphicx}
\begin{document}
\title{Dynamics of hydrogen-like atom bounded by maximal acceleration}
\author{Yaakov Friedman\thanks{Supported in part by German-Israel Foundation for Scientific Research and Development: GIF No. 1078-107.14/2009} and Emanuel Resin\\
Jerusalem College of
Technology\\P.O.B. 16031 Jerusalem 91160, Israel\\email: friedman@jct.ac.il}
\maketitle
\begin{abstract}

The existence of a maximal  acceleration for massive objects was
conjectured by Caianiello  30 years ago based on the Heisenberg uncertainty
relations.  Many consequences of this hypothesis have been studied, but until now, there has been no
evidence that boundedness of the acceleration may lead to quantum behavior.
In previous research, we predicted the existence of a universal maximal acceleration and developed a new dynamics  for which all admissible solutions have an acceleration bounded by the maximal one. Based on  W. K\"{u}ndig's experiment, as reanalyzed  by Kholmetskii et al, we estimated its value to be of the order $10^{19}m/s^2$.

We present here a solution of our dynamical equation for a classical hydrogen-like atom and show
that this dynamics leads to some aspects of quantum behavior. We show that the position of an electron in a hydrogen-like atom can be described only probabilistically. We also show that in this model, the notion of ``center of mass" must be modified. This modification supports  the non-existence of a magnetic moment in the atom and explains the relevance of the conformal group in the quantum region.


\end{abstract}

\section{Introduction}

The existence of a maximal  acceleration for massive objects was
conjectured by Caianiello  \cite{Caianiello} 30 years ago based on the Heisenberg uncertainty
relations. His maximal acceleration was dependent on the mass of the
particle. The consequences of the existence
of a maximal acceleration have been studied by many scientists, and hundreds of papers have been
published in the physics literature,
see for example \cite{scarpetta84}-\cite{Torrome}. The dynamics under the restriction of the
acceleration was studied in \cite{NesterenkoFeoliScarpetta}. The maximal acceleration correction
to the Lamb Shift of Hydrogen, Deuterium and He$^{+}$ was obtained in \cite{LambiasePapiniScarpetta}.
Note, however, that in all these papers, the maximal acceleration is either mass dependent, or, by using Planck's constant, is about $a_m=10^{52}m/s^2$. This value is too large to produce non-classical behavior in the quantum region, where  the accelerations (classically) are much smaller.

In this article, on the other hand, we show that the existence of a maximal acceleration may actually \emph{explain} quantum behavior.
  The existence of a \textit{universal} maximal acceleration was predicted in \cite{FG4} and \cite{FG10} by the following reasoning. It was shown  that from the Generalized Principle of Relativity, it follows that there are only two possible types of transformations between uniformly accelerated systems. If the Clock Hypothesis is not valid, then  there exists a universal maximal acceleration. By \emph{acceleration}, we mean the proper acceleration defined \cite{Rindler} as $\mathbf{g}=d^2\mathbf{x}/dtd\tau$, where $\tau$ is the proper time. This acceleration is equal \cite{FS11} to the four acceleration with respect to the comoving frame. Only this acceleration is bounded, not the acceleration $d^2\mathbf{x}/d\tau^2$. In this case, as shown in \cite{F09} and  \cite{FridmanAnn}, a Doppler-type shift for an accelerated  source will be observed.
The W. K\"{u}ndig experiment \cite{Kundig} tested the transversal Doppler shift of a rotating Mossbauer absorber. This experiment, as reanalyzed \cite{Khoimetski} by Kholmetskii et al, shows that the Doppler shift observed in the experiment differs from the one predicted by Special Relativity. In \cite{F09} we explained that this discrepancy is due to an additional Doppler shift caused by the acceleration, and  we estimated the value of the maximal acceleration $a_m$ to be of the order $10^{19}m/s^2$. A new test for the determination of the maximal acceleration, as described in \cite{Ftest}, is in process.

 Recall that relativistic dynamics stems from the limitation of the speed of moving massive objects by the speed of light. This dynamics reduces to classical mechanics in the case of motion with speed significantly smaller than the speed of light, while it describes accurately the dynamics of objects moving with speed close to the speed of light, which is essentially different from the classical one. We will use here Extended Relativistic Dynamics, introduced in \cite{FridmanAnn}. This new dynamics extends  relativistic dynamics so that all admissible solutions have a speed bounded by the speed of light and an acceleration bounded by the maximal one.
Note that in quantum systems, such as atoms and molecules, the electromagnetic force would generate accelerations
above $10^{19}m/s^2$. Thus, we raise the following question: Can Extended Relativistic Dynamics, which preserves the limitation of the maximal acceleration, also explain quantum phenomena? This paper is the first indication of a positive answer.

 We will calculate the trajectory of the electron in a hydrogen-like atom under
Extended Relativistic Dynamics \cite{FridmanAnn} and show that during a typical time of a quantum measurement, the trajectory of the particle covers a whole area. This may provide an explanation of the probabilistic description of particles in Quantum Mechanics and the uncertainty in the measurement of observable quantities. In addition, our model reveals another non-classical behavior of a hydrogen-like atom. The notion of ``center of mass" must be modified. This leads us to a non-quantum  explanation of the non-existence of a magnetic momentum for a hydrogen atom.

\section{Extended  Relativistic Dynamics equations for hydrogen-like atom}

The main feature of Special Relativity is that the  relativistically
allowed velocities are limited by the speed of light.  Relativistic Dynamics describes motions which preserve this limitation. In Special Relativity, the magnitude of the acceleration is unlimited. In \cite{FridmanAnn} we extended  Relativistic Dynamics  to an \textit{Extended  Relativistic Dynamics }(ERD), a dynamics in which the speed of any moving object is limited  by $c$ and the magnitude of its acceleration is limited by $a_m$.

 In this dynamics we use the 3D position $\mathbf{r}(t)$ and the proper velocity $\mathbf{u}(t)$ to describe the state of a moving object. The proper velocity $\mathbf{u}$ of an object is the derivative of the object's position
with respect to the proper time. In other words, $\mathbf{u}=\gamma \mathbf{v}$, where $\mathbf{v}$ is the object's velocity and $\gamma({v})=1/\sqrt{1-\mathbf{v}^2/c^2}$. In ERD, the evolution of an object of rest-mass $m$ under a 3D force $\mathbf{F}$ is described by the following Hamilton-type system of equations:
\begin{equation}\label{ERD system}
    \left\{
      \begin{array}{l}
       \frac{d\mathbf{r}}{dt} =\frac{\mathbf{u}}{\sqrt{1+|\mathbf{u}|^2/c^2}}\\ \frac{d\mathbf{u}}{dt}=\frac{\mathbf{F}(\mathbf{r},\mathbf{u})/m}
{\sqrt{1+|\mathbf{F}(\mathbf{r},\mathbf{u})|^2/(a_m^2m^2)}}
      \end{array}
    \right .
\end{equation}

For example, if the force $\mathbf{F}$ is a radial one $\mathbf{F}=\mathbf{f}(|\mathbf{r}|)\mathbf{r}/|\mathbf{r}|$, then defining
\[H(\mathbf{r},\mathbf{u})=mc^2\sqrt{1+u^2/c^2}-\int_0^{|\mathbf{r}|}\frac{f(s)ds}{\sqrt{1+f(s)^2/(a_m^2m^2)}}\]
we can rewrite our system (\ref{ERD system}) as
\[\left\{
      \begin{array}{l}
       m\frac{d\mathbf{r}}{dt} =\frac{\partial H}{\partial\mathbf{u}}\\
        m\frac{d\mathbf{u}}{dt}=-\frac{\partial H}{\partial\mathbf{r}}
      \end{array}
    \right .\]

Consider a system of two particles, a proton with mass $m_p=1.7\cdot10^{-27} kg$ and an electron with mass $m_e=9\cdot10^{-31}kg$.
Denote the position and the proper velocity of the proton by $\mathbf{r}_p,\mathbf{u}_p$ and  of the electron  by $\mathbf{r}_e,\mathbf{u}_e$. At this point we will restrict ourselves only to the Coulomb force ignoring the interaction of the particles with the fields. The force of the proton acting on the electron is thus $\mathbf{F_1}=k(\mathbf{r}_p-\mathbf{r}_e)/|\mathbf{r}_p-\mathbf{r}_e|^3$, with $k=2.3\cdot10^{-28}Nm^2$, while the electric force of the electron acting on the proton is $\mathbf{F_2}=k(\mathbf{r}_e-\mathbf{r}_p)/|\mathbf{r}_e-\mathbf{r}_p|^3=-\mathbf{F_1}.$

Typical distances between the proton and the electron are of order $0.5A=0.5\cdot10^{-10}m$. For both the proton and the electron, we have
\begin{equation}\label{Force approx}
   \left| \frac{F_2}{a_m m_p}\right|^2\approx 35 \gg1,\;\;\left| \frac{F_1}{a_m m_e}\right|^2\approx10^8\gg1.
\end{equation}
Thus, the second equation of system (\ref{ERD system}) for both particles becomes
\begin{equation}\label{Erd 2 approx}
   \frac{d\mathbf{u}_p}{dt}\approx a_m\frac{\mathbf{r}_e-\mathbf{r}_p}{|\mathbf{r}_e-\mathbf{r}_p|},\;
   \frac{d\mathbf{u}_e}{dt}\approx a_m\frac{\mathbf{r}_p-\mathbf{r}_e}{|\mathbf{r}_p-\mathbf{r}_e|}\,,
\end{equation}
which implies that the magnitude of the acceleration of each particle will be close to the maximal one.

To estimate the velocities of the particles, we can use the connection of the acceleration and the velocity in circular
motion. Since, in this case, $v^2=aR$, in our system we will have $v\approx10^{4.5}m/s\ll c$. Thus, in the first equation of system
(\ref{ERD system}), we can ignore the denominator, and this equation becomes $\mathbf{u}\approx d\mathbf{r}/dt$. Substituting this into the second equation, we get an approximation of the ERD equation for the particles in a
hydrogen-like atom:
\begin{equation}\label{ERD hydr approx}
  \frac{d^2\mathbf{r}_p}{dt^2}\approx a_m\frac{\mathbf{r}_e-\mathbf{r}_p}{|\mathbf{r}_e-\mathbf{r}_p|},\;
   \frac{d^2\mathbf{r}_e}{dt^2}\approx a_m\frac{\mathbf{r}_p-\mathbf{r}_e}{|\mathbf{r}_p-\mathbf{r}_e|}\,.
\end{equation}

\section{Decomposition into ``center of mass" and their relative motion}

As usual for a two-body problem, in order to solve the system (\ref{ERD hydr approx}) of two particles, we decompose their motion into the motion of
a ``center of mass" and the motion of each particle with respect to the ``center of mass." For our system, the average of these two
equations gives
\begin{equation}\label{cent of mass eqn}
    \frac{d^2(\mathbf{r}_p+\mathbf{r}_e)/2}{dt^2}\approx0.
\end{equation}
This implies that the point $\mathbf{R}:=(\mathbf{r}_p+\mathbf{r}_e)/2$ moves freely, as a ``\textit{center of mass}."
Thus, in this model, the definition of the ``center of mass" differs from the
definition in classical mechanics.

In a classical model for a hydrogen-like atom, the center of mass is positioned at or close to the proton, and the
electron moves around the more or less stationary proton. Hence, classically,  there should be a significant magnetic field for
the atom. In our model, however, both the electron and the proton move around the the new ``center" with similar trajectories. They thus produce
the same magnetic field, but of opposite signs, due to their opposite charges. Thus, the total
magnetic moment of our atom will be almost zero.

We introduce $\mathbf{r}=(\mathbf{r}_p-\mathbf{r}_e)/2$, so that $\mathbf{r}_p=\mathbf{R}+\mathbf{r}$ and
$\mathbf{r}_e=\mathbf{R}-\mathbf{r}$. Subtracting the second equation of (\ref{ERD hydr approx}) from the first and dividing by 2, we obtain
\begin{equation}\label{radial eqn}
    \frac{d^2\mathbf{r}}{dt^2}\approx -a_m\frac{\mathbf{r}}{|\mathbf{r}|},
\end{equation}
which we call the ``radial equation" describing the relative motion of the particles with respect to the center of mass. This is a typical equation for motion in a central field which can be solved by known methods. We will solve this equation by the method of \cite{LLMechanics}, Chapter 14.

\section{Solution of the radial equation}

The solution of equation (\ref{radial eqn}) is in a stable plane generated by the initial vectors $\mathbf{r}(0),\dot{\mathbf{r}}(0)$. Without loss of generality, we may assume that this plane is the $x,y$-plane. We complexify this plane by identifying the point $(x,y)$ with the complex number $\zeta=x+iy=r(t)e^{i\varphi(t)}$.
With this notation, $\frac{\mathbf{r}}{|\mathbf{r}|}=e^{i\varphi(t)}$, and equation (\ref{radial eqn})
becomes
\[\ddot{r}e^{i\varphi}+2i\dot{r}\dot{\varphi}e^{i\varphi}+i\ddot{\varphi}re^{i\varphi}-r\dot{\varphi}^2e^{i\varphi}=-a_me^{i\varphi}.\]
Dividing by $e^{i\varphi}$, we have
\begin{equation}\label{radial plane}
  \ddot{r}+2i\dot{r}\dot{\varphi}+i\ddot{\varphi}r-r\dot{\varphi}^2=-a_m.
\end{equation}

The imaginary part of this equation is
\[2\dot{r}\dot{\varphi}+\ddot{\varphi}r=0\;\;\Rightarrow\;\;\frac{d}{dt}(r^2\dot{\varphi})=0,\]
implying that the angular momentum $r^2\dot{\varphi}$ in the center of mass system is conserved. Denote  the angular momentum,
 defined from the initial conditions, by $c_1$. Then, the equation
 \begin{equation}\label{phi eqn}
  \dot{\varphi}=c_1/r^2 \end{equation}
uniquely defines $\varphi(t)$ if we solve first the equation for $r(t)$ and use the initial conditions.

To find $r(t)$, substitute (\ref{phi eqn}) into the real part of  equation (\ref{radial plane}) to get
$\ddot{r}-\frac{c_1^2}{r^3}+a_m=0\,.$
Multiplying this equation by $2\dot{r}$ and integrating by $t$, we get
    \[ \dot{r}^2=c_2-c_1^2r^{-2}-2a_m r,\]
 with $c_2$ defined by the initial conditions. This equation shows that the radial part can be regarded
 as motion in one dimension in an effective field
 \[ U_{eff}= \frac{c_1^2m}{2r^2}-ma_m r,\]
 where $\frac{c_1^2m}{2r^2}$ is called the centrifugal energy, and $U(r)=ma_m r$ is the potential
 energy.

Since
\begin{equation}\label{r(t) eqn}
  \dot{r}=\pm\sqrt{c_2-c_1^2r^{-2}-2a_m r},
\end{equation}
in order that solutions will exist, the values of $r$ must be restricted by
 \begin{equation}\label{restris radial}
  c_2r^2-c_1^2+2a_m r^3\geq 0.
 \end{equation}
   The solutions of this inequality are $r_1<r< r_2$, where $r_1$ and $r_2$ are the two positive roots of the cubic polynomial in (\ref{restris radial}), which always exist since this polynomial is negative at 0, negative towards $\infty$, and has at least one non-negative value at the initial state.

   Thus, the solutions of the radial equation (\ref{radial eqn}) of  Extended  Relativistic Dynamics for a hydrogen-like atom are obtained by solving  the first-order differential equation (\ref{r(t) eqn}) and then
 (\ref{phi eqn}). It is known that  only for central fields with potential energy proportional to $r^2$ or $1/r$ all finite motions take place in closed paths. The classical electromagnetic field is of this type, but under our dynamics, $U(r)=ma_m r$ is not. Hence, in general, our solution  oscillates between the two radial values $r_1$ and $r_2$ and is not a closed path, see Figure 1.
  \begin{figure}[h!]
  \centering
\scalebox{0.9}{\includegraphics{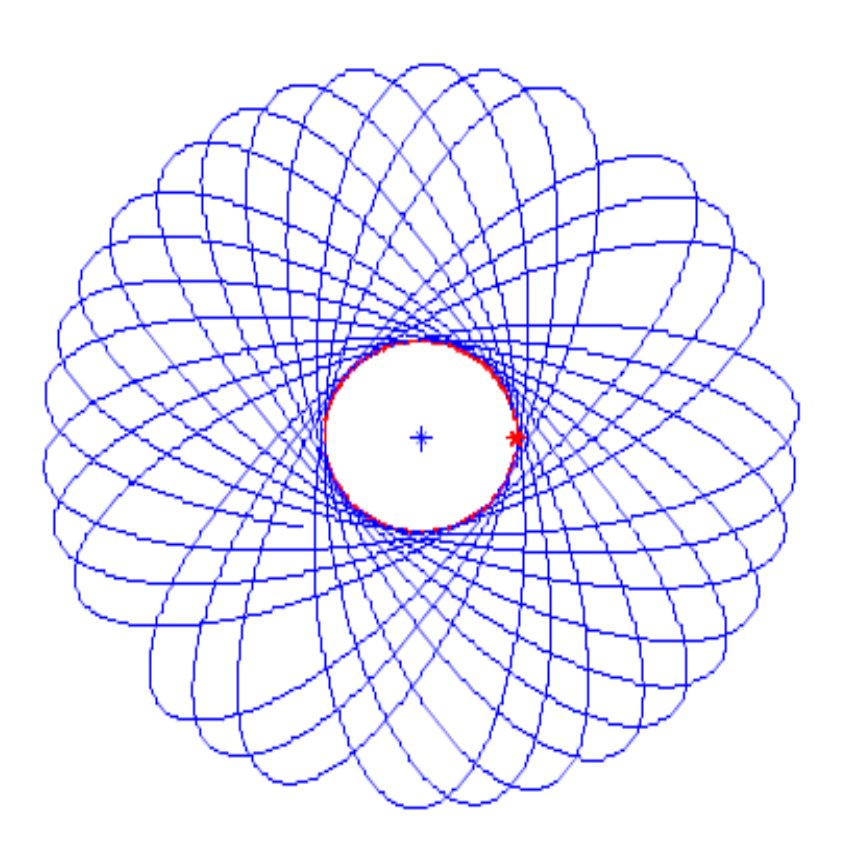}}
  \caption{Electron and proton trajectories in ERD Hydrogen like atom}\label{MaxAccelExp}
\end{figure}
We can get a circular path at the minimum of the effective potential $r=\sqrt[3]{c_1^2/a_m}$, corresponding to the initial velocity $v=\sqrt{ra_m}$, when the velocity is perpendicular to $\mathbf{r}$. For certain discrete values, we can get closed path after $n$ periods.

The frequency of these oscillations can be estimated from the fact that the magnitude of the acceleration is approximately $a_m$, and in approximately circular motion, we have $a=R\omega^2$. These considerations yield a frequency of $\nu\approx10^{14}s^{-1}$. This implies that during one measurement time, the particle will cover a whole area in the annulus $r_1<r<r_2$.

\section{Discussion}

Extended Relativistic Dynamics is an extension of relativistic dynamics in which all admissible solutions have a speed bounded by the speed of light and an acceleration bounded by the maximal acceleration.  Here we have used our estimate $10^{19}m/s^2$ of the uniform maximal acceleration. The existence of a maximal acceleration was also conjectured by Caianiello \cite{Caianiello} based on the Heisenberg  uncertainty relation. We have shown that at quantum system distances, the classical electromagnetic force would generate accelerations above the maximal one. Thus, at the quantum level, Extended Relativistic Dynamics differs significantly from Relativistic Dynamics. Moreover, the existence of a maximal acceleration might explain \textit{why} the behavior of quantum systems differs from the behavior of classical systems.

We have shown that a hydrogen-like atom in Extended Relativistic Dynamics differ significantly from a classical two-body system. We obtained the first approximation of the solution for such system ignoring the interaction of the particles with the field. In a typical time that can be measured, the particle covers a whole area. This may provide an indication of the probabilistic description of particles in Quantum Mechanics.

We have shown that in our model, the expression for the center of mass differs from the classical one. In our model, the total magnetic moment of hydrogen atom is almost zero, which is not so in the classical (non-quantum) model. This observation also reveals the importance of the notion of symmetric velocity, which was introduced in Chapter 2 of \cite{F04}. This velocity is the relativistic half of the regular velocity.  In our model, the velocity of both particles with respect to the new ``center of mass" is the symmetric velocity of the velocity of the electron in the classical model. It is known that the transformations of the symmetric velocities are conformal. Conformal transformations play an important role in the quantum region.

This is only the first step in solving the  hydrogen-like atom by use of Extended Relativistic Dynamics. We
plan to improve our model by: 1. Considering the next approximations of the model. 2. Incorporating the interaction of the charges with the fields. 3. Taking into the consideration the spin of the proton and the electron.

We want to thank Prof. Uziel Sandler and David Hai Gootvilig for helpful remarks and T. Scarr for editorial comments.

  \end{document}